\documentclass[a4paper,fleqn]{cas-sc}

\usepackage[authoryear]{natbib}
\setcitestyle{authoryear, maxcitenames=2, maxbibnames=2}
\usepackage{tcolorbox}
\tcbuselibrary{listingsutf8}
\newtcolorbox[auto counter]{promptbox}[2][]{%
  colback=gray!5!white, 
  colframe=blue!60!teal, 
  fonttitle=\bfseries,
  fontupper=\footnotesize,
  title=Prompt~\thetcbcounter: #2, 
  label={#1}
}
\definecolor{RowGray}{gray}{0.95}
\usepackage{subcaption}
\usepackage{longtable}
\usepackage[table]{xcolor}
\usepackage{array}
\usepackage{tabularx}
\usepackage{xcolor,colortbl}
\usepackage{float}
\usepackage{array}
\usepackage{booktabs}
\usepackage{xcolor}
\usepackage{longtable,booktabs,array}
\usepackage{multirow}
\usepackage{calc} 
\usepackage{stfloats}          
\usepackage[above]{placeins}    
\usepackage{etoolbox}

\makeatletter
\patchcmd\longtable{\par}{\if@noskipsec\mbox{}\fi\par}{}{}
\makeatother
\IfFileExists{footnotehyper.sty}{\usepackage{footnotehyper}}{\usepackage{footnote}}
\makesavenoteenv{longtable}
\ifLuaTeX
  \usepackage{selnolig}  
\fi
\IfFileExists{bookmark.sty}{\usepackage{bookmark}}{\usepackage{hyperref}}
\IfFileExists{xurl.sty}{\usepackage{xurl}}{} 
\hypersetup{
  pdftitle={Latent Profiles of AI Risk Perception and Their Differential Association with Community Driving Safety Concerns: A Person-Centered Analysis},
  hidelinks,
  pdfcreator={LaTeX via pandoc}}
\def\tsc#1{\csdef{#1}{\textsc{\lowercase{#1}}\xspace}}
\tsc{WGM}
\tsc{QE}
\tsc{EP}
\tsc{PMS}
\tsc{BEC}
\tsc{DE}

\begin{document}
\let\WriteBookmarks\relax
\def\floatpagepagefraction{1}
\def\textpagefraction{.001}

\shorttitle{AI Risk Perception and Driving Safety Concerns}

\shortauthors{Rafe et~al.}

\title [mode = title]{Latent Profiles of AI Risk Perception and Their Differential Association with Community Driving Safety Concerns: A Person-Centered Analysis}     

%
\author[1]{Amir Rafe}[orcid=0000-0002-4089-2088]
\cormark[1]
\ead{amir.rafe@txstate.edu}
\credit{Conceptualization, Methodology, Formal analysis, Writing -- original draft, Visualization}

\author[1]{Anika Baitullah}[orcid=0009-0004-9714-0196]
\ead{ffb24@txstate.edu}
\credit{Data curation, Writing -- original draft}

\author[1]{Subasish Das}[orcid=0000-0002-1671-2753]
\ead{subasish@txstate.edu}
\credit{Conceptualization, Methodology, Data curation, Formal analysis, Supervision, Writing -- review \& editing}

\affiliation[1]{organization={Civil Engineering, Texas State University},
    addressline={601 University Drive}, 
    city={San Marcos},
    postcode={78666 TX}, 
    country={USA}}

\cortext[cor1]{Corresponding author}

\begin{abstract}
Public attitudes toward artificial intelligence (AI) and driving safety are typically studied in isolation using variable-centered methods that assume population homogeneity, yet risk perception theory predicts that these evaluations covary within individuals as expressions of underlying worldviews. This study identifies latent profiles of AI risk perception among U.S. adults and tests whether these profiles are differentially associated with community driving-safety concerns. Latent class analysis was applied to nine AI risk-perception indicators from a nationally representative survey (Pew Research Center American Trends Panel Wave 152, $n = 5{,}255$); Bolck--Croon--Hagenaars corrected distal outcome analysis tested class differences on nine driving-safety outcomes, and survey-weighted multinomial logistic regression identified demographic and ideological predictors of class membership. Four classes emerged: Moderate Skeptics (17.5\%), Concerned Pragmatists (42.8\%), AI Ambivalent (10.6\%), and Extreme Alarm (29.1\%), with all nine driving-safety outcomes significantly differentiated across classes. Higher AI concern mapped monotonically onto greater perceived driving-hazard severity; the exception, comparative evaluation of AI versus human driving, was driven by trust rather than concern level. The cross-domain covariation provides person-level evidence for the worldview-based risk structuring posited by Cultural Theory of Risk and yields a four-class segmentation framework for AV communication that links AI risk orientation to transportation safety attitudes.
\end{abstract}



\maketitle

\section{Introduction}


Road traffic crashes constitute one of the most pressing global public health challenges, claiming approximately 1.19 million lives annually and ranking as the leading cause of death among children and young adults aged 5 to 29 \citep{WHO2023}. In the United States, traffic fatalities surged above 40,000 annually during 2021--2023 before declining to an estimated 36,640 in 2025, a level that merely returns the nation to its pre-pandemic 2019 baseline of approximately 36,100 deaths per year \citep{NHTSA2024, NHTSA2026proj}. Even at the projected 2025 level, roughly 100 people die on U.S. roads every day, and human error, encompassing speeding, distracted driving, impaired driving, and aggressive maneuvers, accounts for the vast majority of these crashes. At the same time, artificial intelligence (AI) systems are being embedded in progressively higher-stakes domains, from medical diagnostics and criminal sentencing to vehicle automation \citep{Fagnant2015, SAE2021}. Public attitudes toward both of these developments are consequential: perceptions of AI risk shape regulatory support and technology adoption \citep{Tyson2023}, and perceptions of driving safety influence demand for infrastructure investment and policy intervention \citep{Mannering2016}. Autonomous vehicles (AVs) sit at the precise intersection of these domains, promising to mitigate human-factor crash risk yet simultaneously amplifying public anxieties about algorithmic control, accountability, and systemic failure \citep{Othman2021, Naiseh2025}. Understanding how AI risk perceptions and driving-safety concerns co-occur within individuals is therefore a prerequisite for designing communication strategies that effectively address both domains.


Despite growing scholarly interest in each domain, research on public attitudes toward AI and research on driving-safety perceptions have proceeded largely in isolation. Large-scale surveys document that a majority of U.S. adults express more concern than excitement about AI in daily life, with particular apprehension regarding algorithmic bias, privacy misuse, job displacement, and the spread of inaccurate information \citep{Tyson2023, Dreksler2025}. Separately, an extensive literature examines public acceptance of AVs, focusing on trust, perceived safety, and willingness to ride \citep{Kyriakidis2015, Moody2020, Hilgarter2020, ZhangAV2024}. A critical limitation shared by both bodies of work is their near-exclusive reliance on variable-centered analytic frameworks, principally linear regression and structural equation modeling (SEM), which estimate average associations between predictors and a single outcome across the entire population. These methods implicitly assume that the population is homogeneous in the structure of its attitudes \citep{CollinsLanza2010}. Whether distinct subpopulations exist that organize AI risk perceptions into qualitatively different configurations, and whether such configurations map onto transportation-safety attitudes, remains unknown.


Three streams of prior work are directly relevant to this gap. First, survey research on public attitudes toward AI has established robust demographic gradients: older adults, women, those with lower formal education, and political conservatives tend to express greater concern about AI \citep{Zhang2020, Tyson2023, Dreksler2025}. These findings, however, are derived from models that treat each concern (e.g., bias, privacy, job loss) as a separate dependent variable and each demographic factor as an independent predictor, thereby fragmenting a potentially coherent attitudinal syndrome into isolated bivariate associations. Second, the AV acceptance literature has identified trust, perceived control, and safety expectations as central determinants of willingness to adopt automated driving, yet has examined these constructs without reference to respondents' broader orientations toward AI risk \citep{Bansal2017, Hulse2018, Othman2021, Naiseh2025}. Third, driving-safety perception research has shown that public assessments of community-level hazards, including speeding, impaired driving, and distracted driving, are stratified by demographic characteristics and political orientation, with these factors influencing both the perceived severity of traffic problems and support for enforcement interventions \citep{Ralph2022, MartinezBuelvas2025}. This body of work, however, has not examined whether such assessments covary with broader technology risk attitudes, treating transportation safety as a self-contained perceptual domain. Across all three streams, the analytical strategy presupposes that a single regression coefficient adequately characterizes the population, a premise that theories of risk perception explicitly contest.


The Cultural Theory of Risk \citep{Douglas1982} posits that risk perceptions are not formed in isolation but instead covary across hazard domains because they reflect underlying cultural worldviews. Individuals who perceive one class of technological hazard as threatening tend to extend that apprehension to other domains, a pattern that the psychometric paradigm attributes to shared hazard characteristics such as novelty, controllability, and dread potential \citep{Slovic1987, Dake1991}. AI and driving safety are theoretically linked through these mechanisms: AI is prototypically high in novelty and low in perceived controllability, whereas traffic hazards are familiar yet variable in dread. If risk perceptions cluster into coherent person-level profiles that span both domains, then variable-centered methods, which average over these profiles, will systematically obscure the heterogeneity that is most relevant for targeted policy intervention. Audience segmentation theory \citep{Maibach2011, Leiserowitz2009} reinforces this concern: effective risk communication requires identifying qualitatively distinct audience segments, not formulating messages for an abstract ``average citizen.'' Despite these theoretical predictions, no study has employed person-centered methods to identify latent typologies of AI risk perception and subsequently examined how such typologies predict community-level driving-safety concerns and evaluations of AI versus human driving performance.


The present study addresses this gap by applying latent class analysis (LCA) to nationally representative survey data from the Pew Research Center American Trends Panel (ATP), Wave 152, to identify empirically distinct profiles of AI risk perception among U.S. adults and to evaluate their differential associations with driving-safety attitudes. Three research questions guide the investigation. First, how many empirically distinct latent classes of AI risk-benefit perception exist among U.S. adults, and what are their defining characteristics? Second, do these latent classes differ significantly in their perceptions of community driving-safety problems (speeding, aggressive driving, impaired driving, cellphone distraction, cyclist and pedestrian risk) and in their comparative evaluation of AI versus human driving performance? Third, what demographic and ideological factors (age, gender, education, political ideology, party identification, urbanicity, income, race/ethnicity, and region) predict membership in each latent class?


This study makes four principal contributions to the literature. First, it provides the first person-centered taxonomy of AI risk perception profiles derived from a large national probability sample, moving beyond the variable-centered paradigm that has dominated survey research on public attitudes toward AI. Second, it establishes the first empirical link between AI risk perception profiles and community-level driving-safety concerns by employing the Bolck--Croon--Hagenaars (BCH) method \citep{BolckCroonHagenaars2004, BakkVermunt2016} to relate latent class membership to distal driving-safety outcomes while correcting for classification error, a methodological refinement that is standard in the social and behavioral sciences but has seen virtually no application in transportation research. Third, it integrates three theoretical frameworks, Cultural Theory of Risk, the psychometric paradigm, and audience segmentation theory, into a unified analytic design that explains why AI risk perceptions and driving-safety attitudes should be expected to co-vary at the person level. Fourth, it yields policy-actionable audience segments for AV communication campaigns by identifying which combinations of demographic and ideological characteristics predict membership in each AI risk perception profile.


The remainder of this paper is organized as follows. Section~2 reviews the relevant literature on public AI perception, AV acceptance, driving-safety attitudes, and the theoretical and methodological foundations of person-centered analysis. Section~3 describes the data source, variable operationalization, and the multi-step analytic strategy comprising LCA enumeration, BCH-corrected distal outcome analysis, and multinomial logistic regression with survey weights. Section~4 presents the results, including the four-class latent solution, class-specific driving-safety profiles, and covariate predictors of class membership. Section~5 discusses the theoretical and practical implications of the findings, acknowledges limitations, and outlines directions for future research. Section~6 concludes the paper.

\section{Literature Review}

\subsection{Public Perceptions of Artificial Intelligence}

Survey-based research on public attitudes toward AI has expanded substantially, documenting widespread variation in trust, perceived risk, and acceptance across application domains \citep{Araujo2023, Beets2023, Horowitz2021}. Respondents in the United States and Europe report concern regarding privacy violations, labor displacement, and the erosion of human oversight, even when acknowledging potential benefits of AI in healthcare, transportation, and public administration \citep{Brauner2025, Klein2024, Wei2025}. These patterns are structured by demographic characteristics: age, education, and political orientation are consistently identified as predictors, with older adults, those with lower formal education, and political conservatives reporting higher apprehension \citep{Alasmari2025, Horowitz2021, Yang2025}. Measurement has also evolved from single-item approval indicators toward multidimensional constructs that distinguish among concerns about algorithmic bias, transparency deficits, misinformation, and the displacement of human decision-making authority \citep{Brauner2025, DeSantis2026, Eriksson2026}. Domain-specific risk batteries now permit differentiation between perceptions of AI in healthcare, criminal justice, and transportation contexts \citep{Wei2025, Klein2024}.

Despite these advances in measurement, the analytic strategy across this literature remains almost exclusively variable-centered. The dominant approach regresses a single attitude outcome on demographic or attitudinal predictors, estimating average effects and assuming a continuous, unimodal distribution of attitudes across the population \citep{Araujo2023, Klein2024}. This strategy cannot detect qualitatively distinct subgroups who may organize multiple AI risk concerns into coherent but divergent configurations. If such subgroups exist, as theoretical accounts of risk perception predict, population-level averages will obscure the attitudinal structures most relevant to intervention design.

\subsection{Driving-Safety Perceptions and Autonomous Vehicle Acceptance}

Public perceptions of driving safety encompass a range of community-level hazards, including distracted driving, impaired driving, aggressive behavior, and risks to pedestrians and cyclists \citep{LeeHess2022, MartinezBuelvas2025}. Individuals differ in the severity they attribute to these hazards, with variation linked to personal experience, geographic context, and prior exposure to specific road environments. Perceived risk is consequential for governance: higher concern about traffic risks is associated with greater support for enforcement measures and regulatory action, including automated traffic enforcement and policies protecting vulnerable road users \citep{MartinezBuelvas2025, Ralph2022}. Demographic and ideological characteristics further stratify these perceptions, with political orientation influencing both the perceived severity of traffic hazards and the acceptability of governmental responses \citep{Ralph2022}.

A closely related literature examines public trust in and acceptance of AVs. Acceptance is linked to safety expectations, liability concerns, and willingness to delegate control to algorithmic systems \citep{Hilgarter2020, Ardeshiri2026, Krugel2025, Nazari2026}. Some studies emphasize anticipated benefits such as reductions in human-factor crashes, while others document apprehension about loss of control, algorithmic uncertainty, and accountability gaps \citep{Tennant2025, HoGoh2025}. Political conservatives express lower AV acceptance in U.S. samples, a pattern attributed to broader skepticism toward government-supported technological innovation \citep{Kalambay2026, Mack2021, Peng2020}. Recent work has shifted from framing AV acceptance as an individual adoption decision to examining it as a public safety governance question \citep{Ardeshiri2026, Tennant2025}.

Both bodies of work share two critical limitations. First, they rely on variable-centered methods that estimate average effects and assume population homogeneity, precluding the identification of latent subgroups with distinct perception profiles. Second, they treat driving-safety attitudes and AV evaluations in conceptual isolation from broader AI risk perceptions, despite the theoretical possibility that transportation safety concerns are embedded within more general technology risk worldviews \citep{HoGoh2025, LeeHess2022}.

\subsection{Theoretical Foundations}

Three complementary frameworks ground the expectation that AI risk perceptions and driving-safety attitudes are systematically linked at the person level. Cultural Theory of Risk holds that individuals interpret hazards through stable worldviews that structure perception across domains: a person predisposed to concern about one class of technological hazard is likely to extend that apprehension to adjacent domains \citep{Douglas1982}. Empirical evidence supports this prediction, as attitudes toward AI systems covary with broader ideological and value-based orientations \citep{Araujo2023, Yang2025}. The psychometric paradigm provides a complementary mechanism, explaining perceived risk as a function of hazard characteristics such as controllability, familiarity, and dread potential \citep{Slovic1987}. AI is prototypically high in novelty and low in perceived controllability, attributes associated with elevated risk perception \citep{Klein2024, Wei2025}, whereas driving hazards are more familiar but vary in dread depending on the specific risk. These domain-specific characteristics produce systematic differences in risk evaluation that nonetheless share underlying cognitive dimensions, suggesting that AI and driving-safety perceptions may be linked through common appraisal processes.

Audience segmentation theory shifts the analytic focus to the identification of latent groups with shared attitudinal configurations rather than treating perception as a continuous population-level variable \citep{Maibach2011}. The ``Six Americas'' climate segmentation demonstrated that apparently normal opinion distributions in fact comprise qualitatively distinct segments, each defined by a unique combination of beliefs, risk perceptions, and policy preferences \citep{Leiserowitz2009}. Subsequent applications in risk communication have confirmed that segmentation approaches reveal structures invisible to regression-based methods \citep{Bartolucci2023, PanRyan2024}. Taken together, the three frameworks predict that risk perception across AI and driving-safety domains is structured, multidimensional, and characterized by latent heterogeneity that variable-centered analyses cannot capture.

\subsection{Research Gaps, and Study Positioning}

Person-centered methods such as LCA identify latent subgroups on the basis of response patterns across observed indicators, capturing unobserved heterogeneity by classifying individuals into qualitatively distinct classes \citep{CollinsLanza2010, MagidsonVermunt2004, Masyn2013}. In risk communication, LCA has been applied to identify audience segments in climate engagement \citep{Maibach2011}, public health behavior \citep{Elliott2012}, and disaster risk communication \citep{Bartolucci2023}. Segmentation analysis has also been used to compare AI risk and benefit perceptions between scientists and lay publics \citep{Bao2025}. In transportation, LCA applications remain concentrated on behavioral typologies, including speeding patterns \citep{Peterson2021}, risky driving among novice drivers \citep{Labbo2024}, AV crash patterns \citep{RenXu2025}, mode choice segmentation \citep{PanRyan2024}, and personality-based AV acceptance profiles \citep{Schandl2025}. These studies demonstrate the feasibility of person-centered methods in transportation contexts but focus on behavioral rather than perception-based outcomes.

A further methodological gap concerns the estimation of class-outcome associations. When latent class membership is treated as a known grouping variable, classification error introduces bias into subsequent analyses \citep{BolckCroonHagenaars2004, Vermunt2010}. The Bolck-Croon-Hagenaars (BCH) method addresses this problem by incorporating classification uncertainties into distal outcome estimation, and methodological comparisons demonstrate its superiority over both one-step and naive classify-analyze strategies \citep{BakkVermunt2016, Dziak2016}. Table~\ref{tab:gaps} summarizes the principal gaps this review identifies.

\begin{table}[H]
\caption{Key gaps and implications in AI risk and transportation safety research.}
\label{tab:gaps}
\centering\small
\begin{tabularx}{\textwidth}{>{\raggedright\arraybackslash}p{3.2cm} >{\raggedright\arraybackslash}X >{\raggedright\arraybackslash}p{3.8cm}}
\toprule
\textbf{Gap} & \textbf{Evidence} & \textbf{Consequence} \\
\midrule
Absence of person-centered AI risk typologies & Existing surveys analyze AI perception through variable-centered regression, estimating average predictor effects on single outcomes \citep{Araujo2023, Klein2024}. & Latent audience segments with distinct risk perception configurations cannot be identified. \\
\addlinespace
Limited integration of attitudes toward AI and driving-safety perceptions & AI risk perception and transportation safety attitudes are studied as separate outcomes, with no joint modeling across domains \citep{HoGoh2025, LeeHess2022}. & The cross-domain structure of risk perception remains empirically unexamined. \\
\addlinespace
Absence of BCH correction in transportation research & BCH methods for linking latent classes to distal outcomes have been applied primarily in health and social science contexts \citep{BakkVermunt2016, Dziak2016}. & Transportation studies risk classification-error bias when relating latent classes to external outcomes. \\
\addlinespace
Reliance on demographic predictors as independent effects & AV acceptance studies regress outcomes on individual demographic variables without interaction-based profiling \citep{Mack2021, Nazari2026}. & Cannot identify configurations of ideology and demographics that jointly define distinct audience types. \\
\bottomrule
\end{tabularx}
\end{table}

The present study addresses each of these gaps by applying LCA to nationally representative data to identify distinct profiles of AI risk perception, relating these profiles to community driving-safety concerns through BCH-corrected distal outcome analysis, and examining how demographic and ideological configurations predict profile membership through multinomial logistic regression. This design integrates Cultural Theory of Risk, the psychometric paradigm, and audience segmentation theory within a unified framework to test whether risk perceptions across AI and driving-safety domains reflect coherent, person-level attitudinal structures.

\section{Data and Method}

\begin{figure}
\centering
\includegraphics[width=\columnwidth]{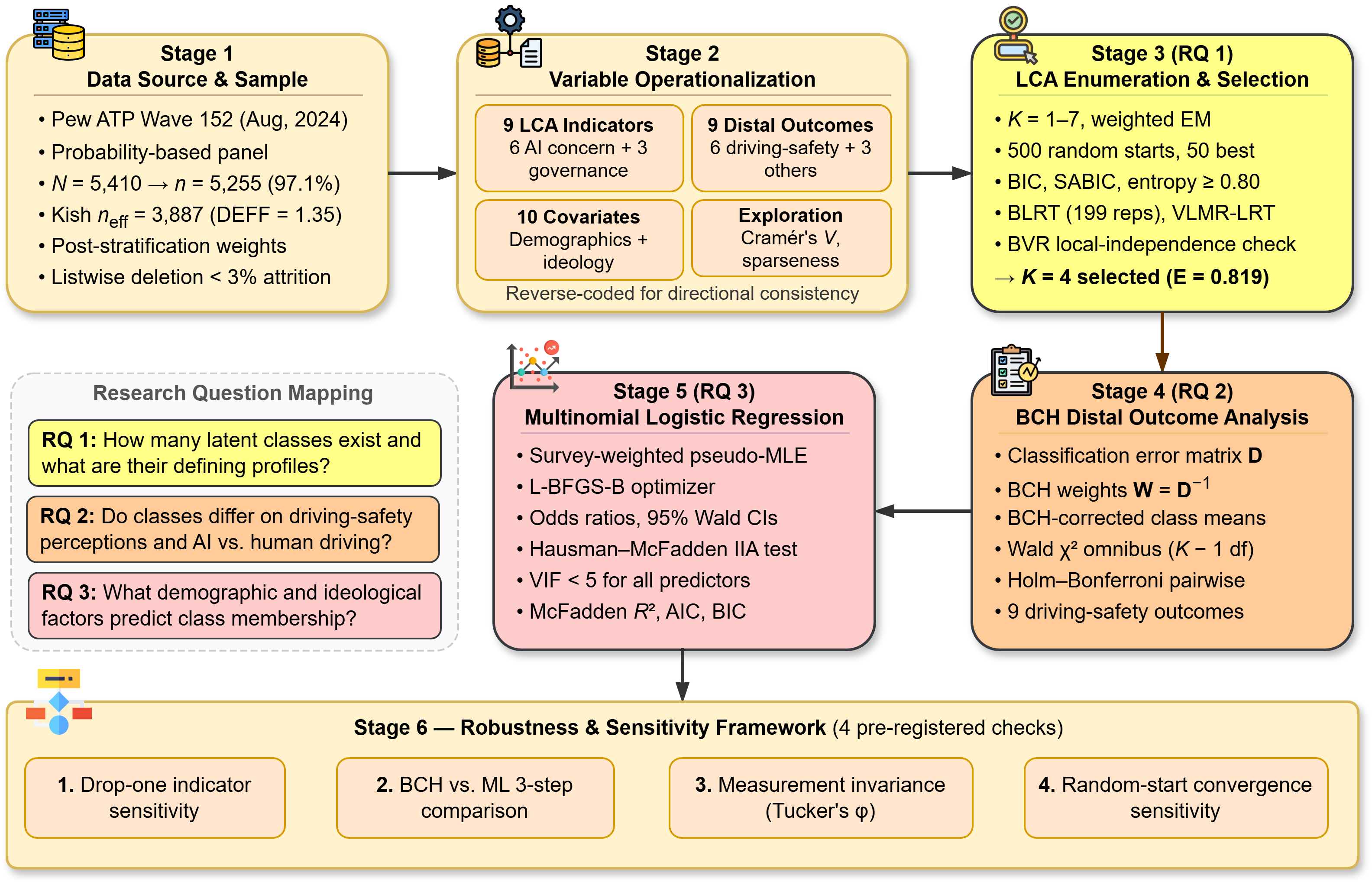}
\caption{Analytic pipeline overview. Six stages proceed from data preparation and variable operationalization through weighted LCA enumeration and model selection (yellow, RQ\,1), BCH-corrected distal outcome analysis (orange, RQ\,2), and survey-weighted multinomial logistic regression (red, RQ\,3), concluding with four robustness checks. The dashed sidebar on the left maps each research question to its corresponding analytic stage.}
\label{fig:pipeline}
\end{figure}

Figure~\ref{fig:pipeline} presents the complete six-stage analytic pipeline that structures this investigation. The remainder of this section describes the data source and sample characteristics, the operationalization of all study variables, and the four-step analytical framework encompassing LCA, BCH-corrected distal outcome testing, multinomial logistic regression, and robustness assessment.

\subsection{Data Source and Sample}

Data were drawn from the ATP Wave 152, a nationally representative online survey of U.S. adults conducted by the Pew Research Center from August 12 through August 18, 2024 \citep{PewATP152}. The ATP is a probability-based panel recruited through address-based sampling, a design that ensures coverage of both internet-connected and non-internet households and avoids the coverage biases inherent in opt-in online panels. Wave 152 included dedicated item batteries on AI risk perception and community driving-safety concerns, making it uniquely suited for joint analysis of these domains within a single nationally representative sample.

The raw sample comprised $N = 5{,}410$ respondents. Listwise deletion was applied exclusively to the nine LCA indicator variables, where item-level missingness ranged from 0\% to 1.04\%, yielding an analytic sample of $n = 5{,}255$ (97.1\% retention). This modest attrition rate falls well below the 5\% threshold at which concerns about missingness-induced bias typically arise \citep{CollinsLanza2010}. Distal outcome variables and covariates retained their original missing values, which were handled per-analysis in subsequent stages: pairwise availability for BCH analyses and listwise deletion for multinomial logistic regression. The maximum covariate missingness was 4.8\% (family income, $n_{\text{missing}} = 260$).

All analyses incorporated the survey-supplied weight \texttt{WEIGHT\_W152}, which adjusts for differential nonresponse and post-stratifies to U.S. Census Bureau population benchmarks on age, gender, education, race/ethnicity, and partisanship. The Kish effective sample size under this weighting scheme was $n_{\text{eff}} = 3{,}887$ with a design effect of 1.35 \citep{Kish1965}, indicating that weighting induced only modest efficiency loss relative to the unweighted sample. The weighted sample closely mirrors the U.S. adult population: 48.8\% male, 50.4\% female; 61.5\% White non-Hispanic, 16.6\% Hispanic, 11.3\% Black non-Hispanic; median age group 30--49; and balanced regional representation across all four Census regions (South 38.4\%, West 23.6\%, Midwest 20.6\%, Northeast 17.5\%).

\subsection{Variable Operationalization}

Table~\ref{tab:variables} presents the complete variable inventory, organized into three panels corresponding to the three analytic roles variables serve in the study design: latent class indicators (Panel A), distal outcome variables (Panel B), and sociodemographic covariates (Panel C).

\begin{table}
\caption{Variable operationalization: latent class indicators, distal outcomes, and covariates.}
\label{tab:variables}
\centering\footnotesize
\begin{tabularx}{\textwidth}{@{} >{\raggedright\arraybackslash}p{3cm} >{\raggedright\arraybackslash}X c >{\raggedright\arraybackslash}p{5.8cm} @{}}
\toprule
\textbf{Variable} & \textbf{Survey Item / Description} & \textbf{Scale} & \textbf{Response Categories} \\
\midrule
\multicolumn{4}{l}{\textit{Panel A: Latent Class Indicators (9 items)}} \\
\midrule
AI concern: Bias & Concern about bias in AI decisions & 1--5 & Not at all $\to$ Extremely concerned \\
AI concern: Impers. & Concern about AI impersonation & 1--5 & Not at all $\to$ Extremely concerned \\
AI concern: Privacy & Misuse of personal information by AI & 1--5 & Not at all $\to$ Extremely concerned \\
AI concern: Inaccuracy & Concern about inaccurate AI information & 1--5 & Not at all $\to$ Extremely concerned \\
AI concern: Connection & Reduced human connection due to AI & 1--5 & Not at all $\to$ Extremely concerned \\
AI concern: Job loss & Concern about job loss from AI & 1--5 & Not at all $\to$ Extremely concerned \\
AI orientation & Excited vs. concerned about AI & 1--3 & More excited / Equal / More concerned \\
AI trust & Trust AI for important decisions & 1--3 & Trust / Not sure / Distrust \\
AI regulation & Views on AI regulation adequacy & 1--3 & Gone too far / Not sure / Not far enough \\
\midrule
\multicolumn{4}{l}{\textit{Panel B: Distal Outcome Variables (9 items)}} \\
\midrule
Safety trend & People driving more or less safely & 1--6 & Much more safely $\to$ Much less safely \\
Speeding & Speeding as a community problem & 1--3 & Not a problem / Minor / Major \\
Aggressive driving & Aggressive driving as a problem & 1--3 & Not a problem / Minor / Major \\
DUI: Alcohol & Alcohol-impaired driving as a problem & 1--3 & Not a problem / Minor / Major \\
DUI: Marijuana & Marijuana-impaired driving as a problem & 1--3 & Not a problem / Minor / Major \\
Cellphone distraction & Cellphone distraction while driving & 1--3 & Not a problem / Minor / Major \\
Cyclist/ped. risk & Risk to cyclists and pedestrians & 1--3 & Not a problem / Minor / Major \\
Road rage freq. & How often respondent experiences road rage & 1--5 & Never $\to$ Very often \\
AI vs. human driving & AI vs. human driving performance & 1--4 & AI much better $\to$ Humans much better \\
\midrule
\multicolumn{4}{l}{\textit{Panel C: Sociodemographic Covariates (10 variables)}} \\
\midrule
Age group & Age bracket & Cat. & 18--29, 30--49, 50--64, 65+ \\
Gender & Self-reported gender & Cat. & Male, Female, Other \\
Education & Highest educational attainment & Ord. & Less than HS $\to$ Postgraduate (6 levels) \\
Ideology & Political ideology & Ord. & Very conservative $\to$ Very liberal (5 levels) \\
Party ID & Party identification & Cat. & Republican/lean, Democrat/lean, Independent \\
Urbanicity & Community type & Cat. & Urban/Suburban, Rural \\
Income & Family income bracket & Ord. & $<$\$30K $\to$ \$150K+ (9 levels) \\
Race/ethnicity & Self-identified race/ethnicity & Cat. & White, Black, Hispanic, Asian, Other \\
Region & U.S. Census region & Cat. & Northeast, Midwest, South, West \\
Driving frequency & How often respondent drives & Ord. & Every day $\to$ Never (6 levels) \\
\bottomrule
\end{tabularx}
\end{table}

Nine items drawn from the ATP Wave 152 battery served as manifest indicators for the LCA measurement model. Six items measured concern about specific AI risks: bias in AI decisions, AI-enabled impersonation, misuse of personal information, inaccurate AI-generated content, reduced human connection, and job displacement. Each was rated on a five-point ordinal scale from ``not at all concerned'' to ``extremely concerned.'' Three additional items captured broader governance orientations toward AI: general excitement versus concern (three categories), trust in AI for important decisions (three categories), and views on the adequacy of current AI regulation (three categories). All items were reverse-coded during data preparation so that higher values uniformly indicate greater concern, skepticism, or demand for regulation, ensuring directional consistency across the indicator set.

Nine driving-safety items from the same survey wave served as distal outcome variables for BCH-corrected analysis. Six items assessed perceived severity of community-level driving hazards: speeding, aggressive driving, alcohol-impaired driving, marijuana-impaired driving, cellphone distraction, and risk to cyclists and pedestrians, each measured on a three-point scale. One item measured the respondent's perception of overall driving-safety trends (six categories), one captured self-reported road rage frequency (five categories), and one evaluated comparative driving performance of AI versus humans (four categories). Critically, these items were excluded from the LCA measurement model and introduced only at the distal outcome analysis stage, preserving the analytic separation required by the BCH framework \citep{BolckCroonHagenaars2004}.

Ten sociodemographic covariates were included as predictors of latent class membership: age group, gender, educational attainment, political ideology, party identification, urbanicity, family income, race/ethnicity, Census region, and personal driving frequency. These covariates were selected on the basis of prior research demonstrating their association with AI attitudes \citep{Tyson2023, Dreksler2025} and driving-safety perceptions \citep{Ralph2022, MartinezBuelvas2025}. Variance inflation factors were computed prior to model estimation to assess multicollinearity.

Prior to model estimation, two exploratory diagnostics were conducted on the nine LCA indicator variables to assess their suitability for latent class extraction. Bias-corrected weighted Cram\'{e}r's $V$ coefficients \citep{Bergsma2013} were computed for all $\binom{9}{2} = 36$ indicator pairs to evaluate bivariate association strength. Joint response pattern sparseness was also quantified by comparing the number of observed unique response patterns to the theoretical cross-classification table.

\subsection{Analytic Strategy}

The analytic strategy comprised four sequential steps, each addressing a specific research question and building on the output of its predecessor (Figure~\ref{fig:pipeline}). All steps incorporated the survey weight \texttt{WEIGHT\_W152}.

\subsubsection{Step 1: Latent class analysis and model selection.}

LCA was employed to identify empirically distinct subgroups of respondents on the basis of their response patterns across the nine AI risk-perception indicators (RQ\,1). Let $\mathbf{Y}_i = (Y_{i1}, Y_{i2}, \dots, Y_{iJ})$ denote the vector of $J = 9$ observed ordinal items for respondent $i$, where each item $Y_{ij}$ has $R_j$ response categories. The model assumes that respondents belong to one of $K$ unobserved latent classes $C_i \in \{1, 2, \dots, K\}$. The marginal likelihood for respondent $i$ is
\begin{equation}
P(\mathbf{Y}_i = \mathbf{y}_i) = \sum_{k=1}^{K} \pi_k \prod_{j=1}^{J} \rho_{j,\, y_{ij},\, k}
\label{eq:lca}
\end{equation}
where $\pi_k = P(C_i = k)$ denotes the class prevalence (mixing proportion), subject to $\sum_{k=1}^{K} \pi_k = 1$ and $\pi_k > 0$, and $\rho_{j,\, y_{ij},\, k} = P(Y_{ij} = y_{ij} \mid C_i = k)$ is the item-response probability for item $j$, response category $y_{ij}$, in class $k$. The defining assumption of the model is \textit{local independence}: conditional on class membership, item responses are mutually independent,
\begin{equation}
P(\mathbf{Y}_i = \mathbf{y}_i \mid C_i = k) = \prod_{j=1}^{J} P(Y_{ij} = y_{ij} \mid C_i = k)
\label{eq:locind}
\end{equation}

Parameters $\{\pi_k, \rho_{j,r,k}\}$ were estimated by maximizing the survey-weighted log-likelihood via the expectation-maximization (EM) algorithm \citep{DempsterLairdRubin1977}:
\begin{equation}
\ell(\boldsymbol{\theta}) = \sum_{i=1}^{n} w_i \log \left[ \sum_{k=1}^{K} \pi_k \prod_{j=1}^{J} \rho_{j,\, y_{ij},\, k} \right]
\label{eq:loglik}
\end{equation}
where $w_i$ is the survey weight for respondent $i$. To mitigate sensitivity to local maxima, 500 random starting values were generated and the 50 best solutions (ranked by log-likelihood) were carried to full convergence (tolerance $= 10^{-8}$).

Models for $K = 1$ through $K = 7$ classes were estimated and compared using a triangulation of statistical and substantive criteria. The Bayesian information criterion \citep{Schwarz1978} and sample-size adjusted BIC (SABIC) were computed as
\begin{equation}
\text{BIC} = -2\ell(\hat{\boldsymbol{\theta}}) + p \ln(n), \qquad
\text{SABIC} = -2\ell(\hat{\boldsymbol{\theta}}) + p \ln\!\left(\frac{n+2}{24}\right)
\label{eq:bic}
\end{equation}
where $p$ is the number of freely estimated parameters and $n$ the sample size. Classification accuracy was assessed via entropy,
\begin{equation}
E_K = 1 - \frac{-\sum_{i=1}^{n} \sum_{k=1}^{K} \hat{p}_{ik} \ln \hat{p}_{ik}}{n \ln K}
\label{eq:entropy}
\end{equation}
where $\hat{p}_{ik} = P(C_i = k \mid \mathbf{Y}_i, \hat{\boldsymbol{\theta}})$ is the posterior class membership probability. Values of $E_K \geq 0.80$ are conventionally interpreted as indicating clear classification \citep{Masyn2013}. Two likelihood-ratio tests were employed to compare adjacent models: the parametric bootstrap likelihood-ratio test (BLRT; 199 replicates with uniform weights on simulated data) and the Lo--Mendell--Rubin adjusted likelihood-ratio test (VLMR-LRT) \citep{LoMendellRubin2001}. Significance ($p < .05$) on either test supports the retention of the additional class. Final model selection also required that all classes exceed 5\% of the sample and that the resulting profiles be substantively interpretable and theoretically coherent \citep{Nylund2007}. Adherence to the local independence assumption was evaluated using bivariate residual (BVR) statistics for all $\binom{J}{2} = 36$ item pairs under the selected model \citep{MagidsonVermunt2004}.

\subsubsection{Step 2: BCH-corrected distal outcome analysis.}

To address RQ\,2, class differences on the nine driving-safety distal outcomes were tested using the BCH method \citep{BolckCroonHagenaars2004}, a three-step procedure that corrects for classification error when relating latent classes to external variables. The correction is necessary because modal class assignment (assigning each respondent to their highest-posterior class) introduces misclassification bias that systematically attenuates between-class differences \citep{BakkVermunt2016}. The procedure is as follows.

From the LCA solution, the classification error matrix $\mathbf{D}$ is computed from the posterior probabilities, with entries
\begin{equation}
D_{km} = P(\hat{C}_i = m \mid C_i = k) = \frac{1}{N_k} \sum_{i:\, \hat{C}_i = k} \hat{p}_{im}
\label{eq:dmatrix}
\end{equation}
where $\hat{C}_i$ is the modal class assignment and $N_k$ is the number of respondents assigned to class $k$. The BCH weight matrix is obtained by direct inversion, $\mathbf{W} = \mathbf{D}^{-1}$. For each distal outcome variable $Z$, BCH-corrected class-specific means are then estimated as
\begin{equation}
\hat{\mu}_{k}^{\text{BCH}} = \frac{\sum_{i=1}^{n} w_i \, W_{k,\hat{C}_i} \, Z_i}{\sum_{i=1}^{n} w_i \, W_{k,\hat{C}_i}}
\label{eq:bchmean}
\end{equation}

Between-class equality is tested via a Wald $\chi^2$ statistic with $K - 1$ degrees of freedom:
\begin{equation}
\chi^2_{\text{Wald}} = \left(\hat{\boldsymbol{\mu}}^{\text{BCH}}\right)^{\!\top} \mathbf{C}^{\top} \left(\mathbf{C} \, \hat{\boldsymbol{\Sigma}} \, \mathbf{C}^{\top}\right)^{-1} \mathbf{C} \, \hat{\boldsymbol{\mu}}^{\text{BCH}}
\label{eq:wald}
\end{equation}
where $\mathbf{C}$ is a contrast matrix encoding $H_0\!: \mu_1 = \mu_2 = \cdots = \mu_K$ and $\hat{\boldsymbol{\Sigma}}$ is the sandwich-estimated covariance matrix of the BCH-corrected means. Pairwise post-hoc comparisons among all $\binom{K}{2}$ class pairs were conducted with the Holm--Bonferroni sequential correction \citep{Holm1979} to control the family-wise error rate within each outcome. The analytic separation of the LCA measurement model (Step 1) from the distal outcome step (Step 2) ensures that driving-safety items do not influence class formation, satisfying the fundamental assumption of the BCH framework.

\subsubsection{Step 3: Multinomial logistic regression.}

To address RQ\,3, the probability of belonging to class $k$ relative to a reference class was modeled as a function of the ten sociodemographic covariates $\mathbf{x}_i$ via survey-weighted multinomial logistic regression:
\begin{equation}
\log \frac{P(C_i = k \mid \mathbf{x}_i)}{P(C_i = K \mid \mathbf{x}_i)} = \alpha_k + \boldsymbol{\beta}_k^{\top} \mathbf{x}_i, \qquad k = 1, \dots, K-1
\label{eq:mnl}
\end{equation}

The model was estimated using pseudo-maximum-likelihood with the L-BFGS-B optimizer, incorporating the survey weight $w_i$ throughout. Exponentiated coefficients yield odds ratios with 95\% Wald confidence intervals:
\begin{equation}
\text{OR}_{k,\ell} = \exp(\hat{\beta}_{k,\ell}), \qquad
\text{CI}_{95\%} = \exp\!\left(\hat{\beta}_{k,\ell} \pm 1.96 \, \widehat{\text{SE}}(\hat{\beta}_{k,\ell})\right)
\label{eq:or}
\end{equation}

Three diagnostic checks were performed to verify model assumptions. The independence of irrelevant alternatives (IIA) was tested via the Hausman--McFadden test \citep{HausmanMcFadden1984}. Multicollinearity among covariates was assessed via variance inflation factors (VIF $< 5$ for all predictors). Overall model fit was evaluated using the McFadden pseudo-$R^2$, AIC, and BIC. Predicted class membership probabilities were computed for theoretically motivated covariate profiles to illustrate how specific combinations of demographic and ideological characteristics translate into differential class membership.

\subsubsection{Step 4: Robustness checks.}

Four robustness checks were conducted to assess the stability and generalizability of the latent class solution. First, indicator sensitivity was evaluated by fitting nine leave-one-out LCA models, each omitting one of the nine indicators, and assessing whether the four-class structure was recoverable with all class prevalence shifts below 0.10. Second, the equivalence of the BCH estimator and the maximum-likelihood three-step estimator \citep{Vermunt2010} was confirmed for continuous distal outcomes, and the magnitude of attenuation bias introduced by naive modal-assignment analysis was quantified. Third, configural measurement invariance was assessed across political subgroups (Republican versus Democrat) by estimating separate $K = 4$ models for each subsample and computing Tucker's congruence coefficient $\phi$ \citep{LorenzoSevaTenBerge2006}, with $\phi > 0.85$ indicating acceptable profile similarity across groups. Fourth, random-start sensitivity was evaluated by re-estimating the $K = 4$ model with 50, 100, 200, 500, and 1{,}000 random starting values to confirm convergence to a unique global optimum.

\section{Results}

\subsection{Exploratory Indicator Diagnostics}

Table~\ref{tab:cramersv} presents the bias-corrected weighted Cram\'{e}r's $V$ matrix for all 36 pairwise associations among the nine LCA indicators. All associations were statistically significant ($p < 0.001$). The six domain-specific concern items formed a block of moderate intercorrelations ($V = 0.327$--$0.447$), with the strongest association between privacy and inaccuracy ($V = 0.447$). The three governance orientation items showed weaker associations with the concern block ($V = 0.088$--$0.271$), consistent with their distinct response formats. This pattern indicates that the indicators share sufficient common variance to support latent class extraction while retaining enough unique information to differentiate profiles.

\begin{table}[H]
\caption{Bias-corrected weighted Cram\'{e}r's $V$ among the nine LCA indicators ($n = 5{,}255$).}
\label{tab:cramersv}
\centering\footnotesize
\begin{tabular}{@{} l c c c c c c c c c @{}}
\toprule
 & Bias & Impers. & Privacy & Inacc. & Connect. & Job loss & Orient. & Trust & Regul. \\
\midrule
Bias         & ---   &       &       &       &       &       &       &       &       \\
Impers.      & 0.327  & ---   &       &       &       &       &       &       &       \\
Privacy      & 0.399  & 0.418  & ---   &       &       &       &       &       &       \\
Inacc.       & 0.410  & 0.412  & 0.447  & ---   &       &       &       &       &       \\
Connect.     & 0.335  & 0.351  & 0.347  & 0.351  & ---   &       &       &       &       \\
Job loss     & 0.327  & 0.346  & 0.381  & 0.335  & 0.384  & ---   &       &       &       \\
Orient.      & 0.203  & 0.200  & 0.243  & 0.231  & 0.225  & 0.271  & ---   &       &       \\
Trust        & 0.158  & 0.134  & 0.170  & 0.176  & 0.177  & 0.152  & 0.233  & ---   &       \\
Regul.       & 0.088  & 0.168  & 0.134  & 0.134  & 0.117  & 0.114  & 0.139  & 0.160  & ---   \\
\bottomrule
\end{tabular}
\begin{flushleft}
\scriptsize\textit{Note.} All 36 coefficients significant at $p < 0.001$. Impers.\ = Impersonation; Inacc.\ = Inaccuracy; Connect.\ = Human connection; Orient.\ = Excited vs.\ concerned; Regul.\ = Regulation.
\end{flushleft}
\end{table}

Joint response pattern sparseness was high: the nine polytomous indicators generate a theoretical cross-classification table of $5^6 \times 3^3 = 421{,}875$ cells, of which only 3,197 (0.76\%) were observed in the analytic sample, yielding an overall sparseness ratio of 0.992. This degree of sparseness is typical of polytomous ordinal LCA and does not impede maximum-likelihood estimation, which operates on marginal and pairwise sufficient statistics rather than the full contingency table \citep{CollinsLanza2010}.

\subsection{Model Enumeration and Selection}

\begin{figure}
\centering
\includegraphics[width=\columnwidth]{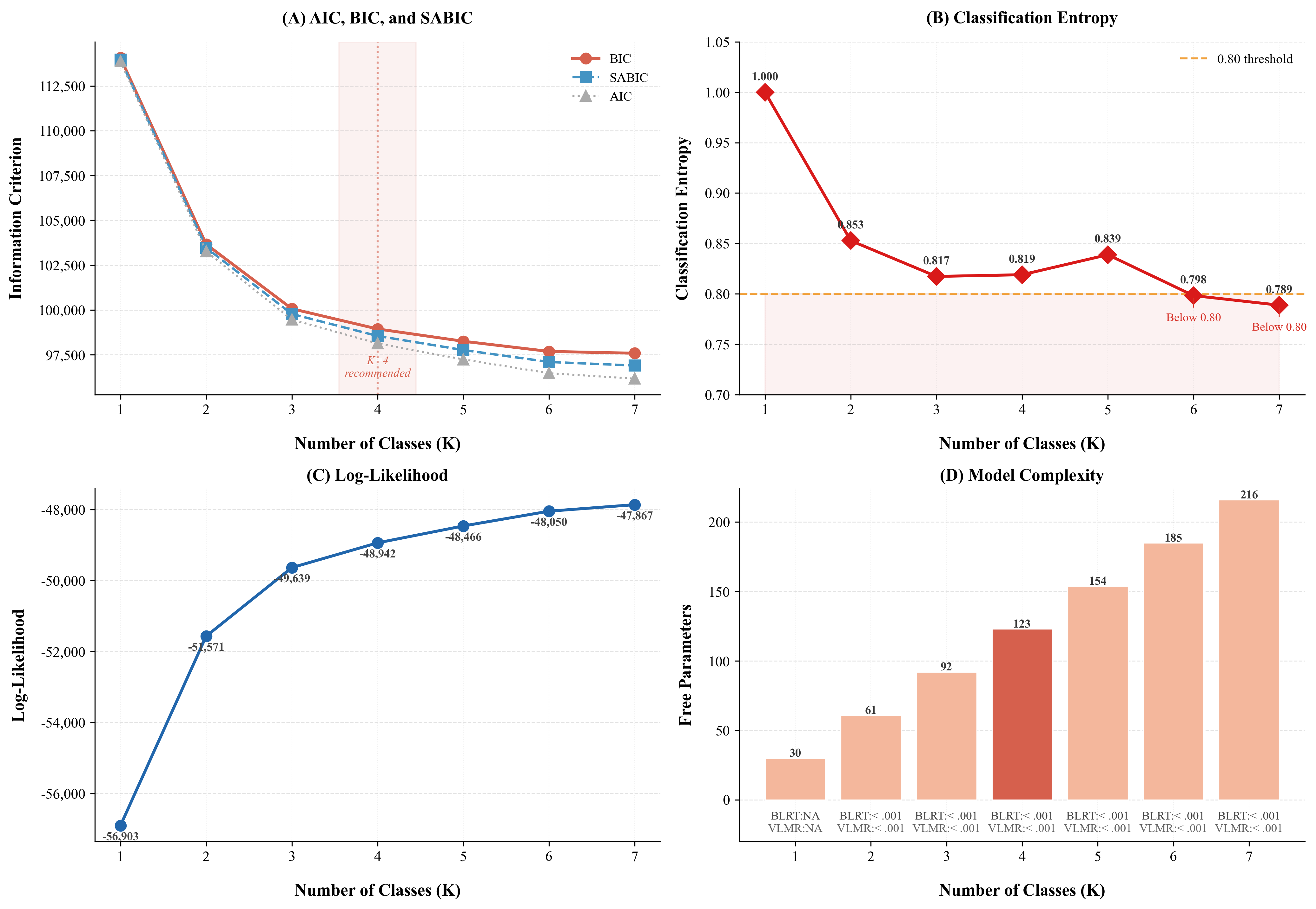}
\caption{Model fit indices for latent class solutions ($K = 1$--$7$). (A) BIC, SABIC, and AIC with the shaded region highlighting the $K = 4$ elbow. (B) Classification entropy with the 0.80 adequacy threshold. (C) Log-likelihood trajectory. (D) Free parameters per model with BLRT and VLMR-LRT significance annotations. The red bar marks the recommended $K = 4$ solution.}
\label{fig:fit}
\end{figure}

Table~\ref{tab:fit} presents the fit indices for $K = 1$ through $K = 7$ latent class models, and Figure~\ref{fig:fit} plots the BIC, entropy, and likelihood trajectories to assist visual identification of the optimal solution. The BIC decreased monotonically from 114,064 ($K = 1$) to 97,584 ($K = 7$), with no global minimum within the tested range; however, the marginal improvement diminished substantially beyond $K = 4$. The BIC decrement from $K = 3$ to $K = 4$ was 1,128 points, compared to 686 points from $K = 4$ to $K = 5$, representing a 39\% reduction in improvement. The SABIC exhibited a parallel elbow pattern. Entropy remained above the 0.80 threshold for $K \leq 5$ (range: 0.817 to 0.853) and fell below it at $K = 6$ (0.798). Both the BLRT and VLMR-LRT were significant ($p < 0.001$) at every step from $K = 2$ through $K = 7$, indicating that each additional class improved model fit in absolute terms. However, the $K = 5$ solution was rejected on substantive grounds: its smallest class contained only 1.5\% of the sample, well below the 5\% minimum recommended by \citet{Nylund2007}. The $K = 4$ solution satisfied all triangulated criteria: a clear BIC elbow, entropy of 0.819, significant sequential tests, a smallest class of 10.9\%, and four substantively interpretable profiles. This solution was retained for all subsequent analyses.

\begin{table}[H]
\caption{Latent class enumeration: model fit indices ($n = 5{,}255$).}
\label{tab:fit}
\centering
\begin{tabular}{@{} c r c r r r c c c @{}}
\toprule
$K$ & LL & $n_{\text{par}}$ & BIC & SABIC & Entropy & BLRT $p$ & VLMR $p$ \\
\midrule
1 & $-$56,904 & 30  & 114,064 & 113,969 & 1.000 & ---   & ---   \\
2 & $-$51,571 & 61  & 103,665 & 103,471 & 0.853 & $<$0.001 & $<$0.001 \\
3 & $-$49,639 & 92  & 100,067 & 99,774  & 0.817 & $<$0.001 & $<$0.001 \\
\textbf{4} & $\mathbf{-}$\textbf{48,942} & \textbf{123} & \textbf{98,939} & \textbf{98,548} & \textbf{0.819} & $<$\textbf{0.001} & $<$\textbf{0.001} \\
5 & $-$48,467 & 154 & 98,252  & 97,763  & 0.839 & $<$0.001 & $<$0.001 \\
6 & $-$48,050 & 185 & 97,686  & 97,098  & 0.798 & $<$0.001 & $<$0.001 \\
7 & $-$47,867 & 216 & 97,584  & 96,898  & 0.789 & $<$0.001 & $<$0.001 \\
\bottomrule
\end{tabular}
\begin{flushleft}
\scriptsize\textit{Note.} LL = log-likelihood; $n_{\text{par}}$ = free parameters; BIC = Bayesian information criterion; SABIC = sample-size adjusted BIC. Weighted EM with 500 random starts, 50 best to convergence. BLRT = bootstrap likelihood-ratio test (199 replicates). VLMR = Lo--Mendell--Rubin adjusted LRT. Bold row indicates the selected model.
\end{flushleft}
\end{table}

\subsection{Class Profiles and Interpretation}

\begin{figure}
\centering
\includegraphics[width=\columnwidth]{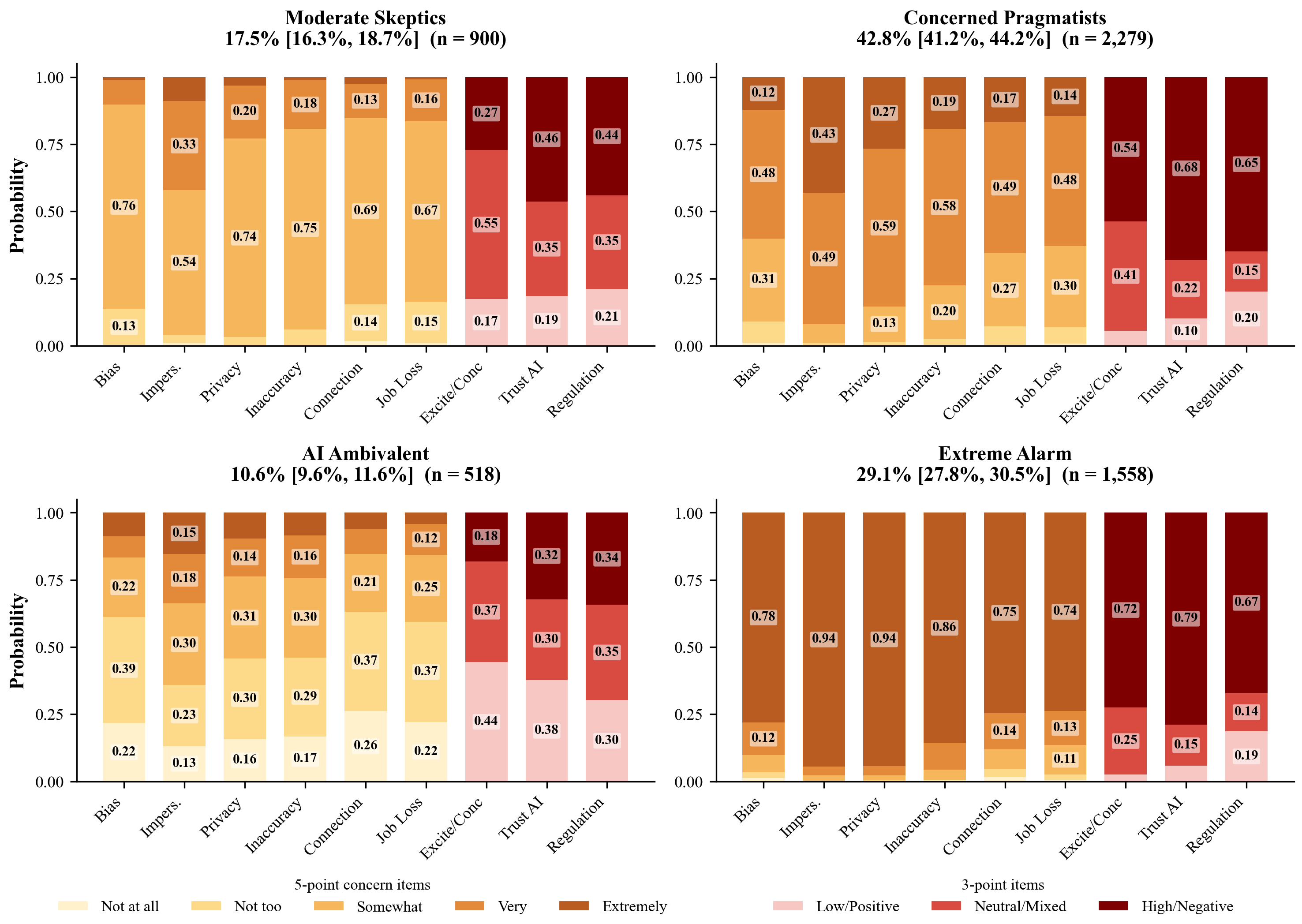}
\caption{Item-response probability profiles for the four-class solution ($K = 4$). Each panel displays one of the nine LCA indicator items; within each panel, the four classes are shown with response-category probabilities. Class labels and weighted prevalences: C1 = Moderate Skeptics (17.5\%), C2 = Concerned Pragmatists (42.8\%), C3 = AI Ambivalent (10.6\%), C4 = Extreme Alarm (29.1\%).}
\label{fig:irp}
\end{figure}

Figure~\ref{fig:irp} displays the item-response probability profiles for the four-class solution. The weighted class prevalences with 95\% bootstrap confidence intervals are as follows: Class~1 (Moderate Skeptics), 17.5\% [16.3, 18.7]; Class~2 (Concerned Pragmatists), 42.8\% [41.2, 44.2]; Class~3 (AI Ambivalent), 10.6\% [9.6, 11.6]; and Class~4 (Extreme Alarm), 29.1\% [27.8, 30.5].

Class~1 (Moderate Skeptics) is characterized by modal ``somewhat concerned'' responses across all six AI concern items, with item-response probabilities for this category ranging from 0.538 to 0.762. Members of this class are roughly evenly divided between distrusting and being unsure about AI (probabilities of 0.463 and 0.352, respectively) and are split between viewing regulation as insufficient (0.440) and being unsure (0.348). This profile reflects a moderate, noncommittal orientation toward AI risks: respondents acknowledge concerns but do not endorse them strongly.

Class~2 (Concerned Pragmatists), the largest class comprising nearly half the sample, exhibits ``very concerned'' as the dominant response category on all six concern items (probabilities 0.480 to 0.586). A clear majority distrust AI for important decisions (0.679) and believe that regulation has not gone far enough (0.648). Members are more concerned than excited about AI (0.536) but do not reach the ceiling levels of alarm observed in Class~4. This class represents the pragmatic mainstream: its members treat AI risks as serious and favor greater oversight, but their concern is tempered relative to extreme levels.

Class~3 (AI Ambivalent), the smallest class, is distinguished by highly dispersed response probabilities across concern levels; no single category dominates, and the ``not at all'' and ``not too concerned'' categories receive combined probabilities between 0.46 and 0.63 across items. This class has the highest proportion endorsing trust in AI (0.377), the greatest excitement about AI (0.444 ``more excited''), and the most evenly divided views on regulatory adequacy. The label ``ambivalent'' reflects genuine uncertainty rather than low engagement: these respondents express neither strong concern nor unqualified enthusiasm and appear open to multiple evaluative positions.

Class~4 (Extreme Alarm) is defined by overwhelming endorsement of ``extremely concerned'' across all six concern items, with probabilities ranging from 0.738 (job loss) to 0.944 (AI impersonation and privacy misuse). This class exhibits the highest distrust of AI (0.789), the strongest concern orientation (0.725 ``more concerned than excited''), and strong support for additional regulation (0.671). The profile represents a comprehensive, internally consistent pattern of maximal AI concern that extends uniformly across all assessed risk domains.

Classification quality for the four-class solution met conventional benchmarks. The average diagonal entry of the classification error matrix was 0.893, with individual class accuracies ranging from 0.858 (Concerned Pragmatists) to 0.950 (AI Ambivalent), all exceeding the 0.80 threshold for reliable class assignment \citep{Masyn2013}. Evaluation of local independence revealed BVR violations in 26 of 36 indicator pairs. These violations formed two coherent blocks: one comprising the six domain-specific concern items, which share a common five-point response format, and another comprising the three governance orientation items, whose response formats differ from the concern items. Residual within-class association among items sharing identical response scales is a documented property of polytomous ordinal LCA models \citep{MagidsonVermunt2004} and does not invalidate the class solution when the violation pattern is structurally interpretable, as it is here.

\subsection{Driving-Safety Outcomes: BCH Analysis}

Table~\ref{tab:bch} reports the BCH-corrected class-specific means and omnibus Wald $\chi^2$ tests for all nine distal driving-safety outcomes. All nine omnibus tests were statistically significant, confirming that latent class membership is associated with differential driving-safety perceptions across every assessed outcome. The strongest class differentiation was observed for cellphone distraction ($\chi^2(3) = 55.68$, $p < 0.001$), road rage frequency ($\chi^2(3) = 46.50$, $p < 0.001$), and aggressive driving ($\chi^2(3) = 45.24$, $p < 0.001$). The weakest, though still significant, differentiation was observed for marijuana-impaired driving ($\chi^2(3) = 11.60$, $p = 0.009$) and AI versus human driving ($\chi^2(3) = 13.04$, $p = 0.005$).

\begin{table}[H]
\caption{BCH-corrected class-specific means and omnibus Wald $\chi^2$ tests for distal outcomes ($K = 4$).}
\label{tab:bch}
\centering
\begin{tabular}{@{} l c c c c r c @{}}
\toprule
 & C1: Mod. & C2: Conc. & C3: AI & C4: Ext. & & \\
Outcome & Skeptics & Pragmatists & Ambivalent & Alarm & $\chi^2(3)$ & $p$ \\
\midrule
Safety trend (1--6) & 3.65 (0.07) & 3.78 (0.03) & 3.76 (0.11) & 3.99 (0.04) & 29.80 & $<$0.001 \\
Speeding (1--3) & 1.55 (0.03) & 1.39 (0.02) & 1.59 (0.05) & 1.38 (0.02) & 33.73 & $<$0.001 \\
Aggress. driving (1--3) & 1.61 (0.04) & 1.41 (0.02) & 1.55 (0.05) & 1.35 (0.02) & 45.24 & $<$0.001 \\
DUI alcohol (1--3) & 1.68 (0.04) & 1.54 (0.02) & 1.68 (0.06) & 1.49 (0.02) & 26.93 & $<$0.001 \\
DUI marijuana (1--3) & 1.87 (0.04) & 1.78 (0.02) & 1.90 (0.06) & 1.74 (0.02) & 11.60 & 0.009 \\
Cellphone distr. (1--3) & 1.39 (0.03) & 1.20 (0.01) & 1.45 (0.05) & 1.19 (0.02) & 55.68 & $<$0.001 \\
Cyclist/ped. risk (1--3) & 1.74 (0.04) & 1.65 (0.02) & 1.79 (0.06) & 1.54 (0.02) & 33.74 & $<$0.001 \\
Road rage (1--5) & 2.88 (0.05) & 3.00 (0.02) & 2.91 (0.08) & 3.22 (0.03) & 46.50 & $<$0.001 \\
AI vs. human (1--4) & 2.64 (0.06) & 2.47 (0.03) & 2.32 (0.09) & 2.42 (0.03) & 13.04 & 0.005 \\
\bottomrule
\end{tabular}
\begin{flushleft}
\scriptsize\textit{Note.} Standard errors in parentheses. Means estimated via the BCH method \citep{BolckCroonHagenaars2004} with survey weights. For driving-problem items (1--3 scale), lower values indicate greater perceived severity. For safety trend, higher values indicate more pessimistic assessment. For road rage, higher values indicate greater frequency. For AI vs.\ human driving, higher values indicate more favorable evaluation of human driving.
\end{flushleft}
\end{table}

\begin{figure}
\centering
\includegraphics[width=\columnwidth]{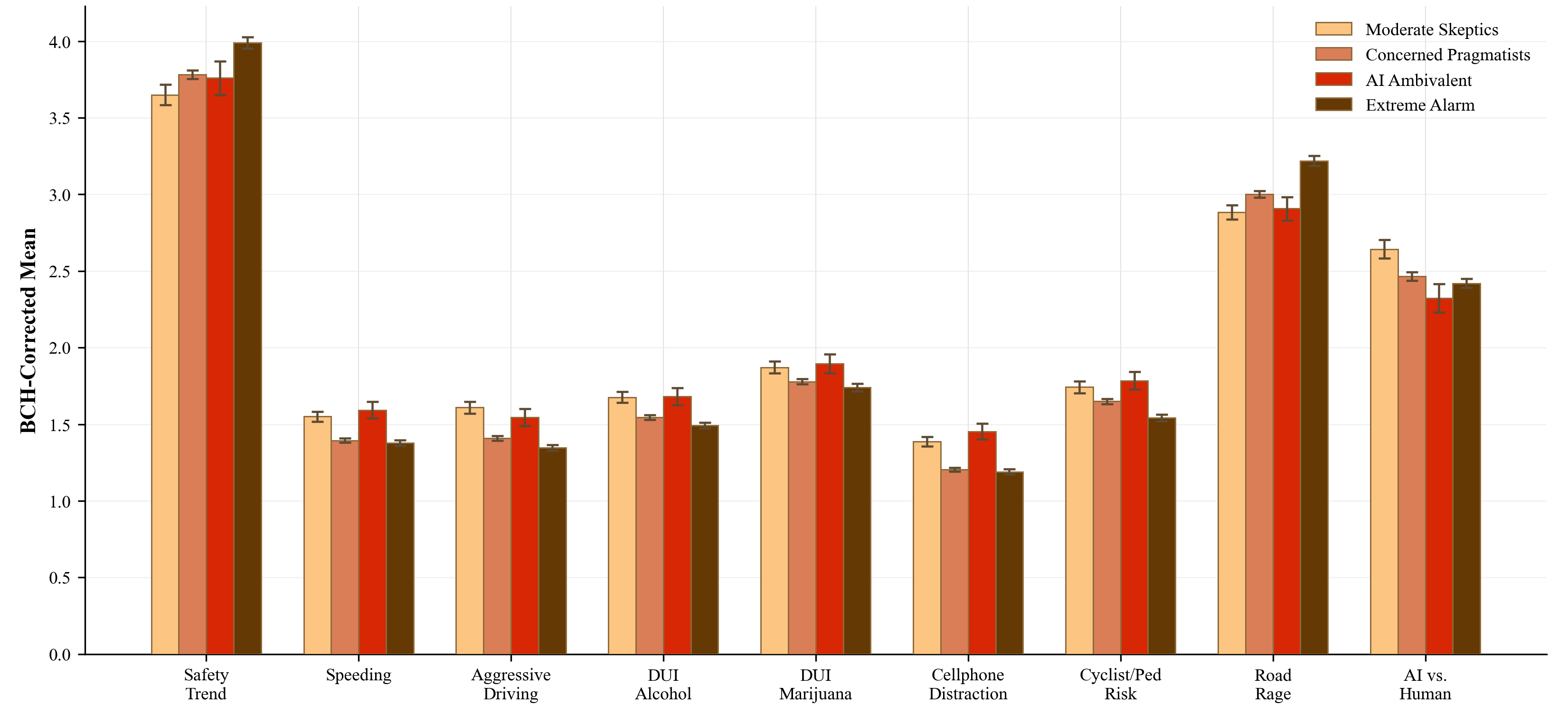}
\caption{BCH-corrected class-specific means for nine driving-safety distal outcomes. Error bars denote sandwich-estimated standard errors. For driving-problem items (1--3 scale), lower bars indicate greater perceived severity. For road rage (1--5) and safety trend (1--6), higher bars indicate more pessimistic assessment and greater frequency, respectively.}
\label{fig:bch}
\end{figure}

Across the six community driving-problem items (speeding, aggressive driving, DUI alcohol, DUI marijuana, cellphone distraction, and cyclist/pedestrian risk), a consistent ordering emerged (Figure~\ref{fig:bch}). Class~4 (Extreme Alarm) perceived all six hazards as most severe, followed by Class~2 (Concerned Pragmatists), while Classes~1 (Moderate Skeptics) and~3 (AI Ambivalent) perceived driving problems as least severe. This monotonic pattern, in which greater AI risk concern maps onto greater driving-safety concern, is consistent with the cross-domain risk hypothesis derived from Cultural Theory \citep{Douglas1982}. For the driving-safety trend item, Class~4 reported the most pessimistic assessment ($M = 3.99$), perceiving a trend toward less safe driving, whereas Class~1 reported the most moderate assessment ($M = 3.65$). Road rage frequency followed the same pattern: Class~4 reported the highest frequency ($M = 3.22$), significantly exceeding all other classes.

The AI versus human driving comparison produced a distinctive pattern that departs from the general severity ordering. Class~1 (Moderate Skeptics), not Class~4 (Extreme Alarm), expressed the most skeptical evaluation of AI driving relative to human driving ($M = 2.64$), while Class~3 (AI Ambivalent) evaluated AI driving most favorably ($M = 2.32$). Classes~2 and~4, despite their markedly different levels of AI concern, converged on intermediate evaluations ($M = 2.47$ and $M = 2.42$, respectively). This result indicates that the comparative evaluation of AI versus human driving competence is not simply a function of general AI concern but appears to be more closely related to AI trust: Class~1 combines moderate concern with the second-highest distrust of AI (0.463), whereas Class~3, with the lowest concern, has the highest trust (0.377). Pairwise comparisons with Holm--Bonferroni correction confirmed that DUI marijuana was the most weakly discriminating outcome, with only one significant pairwise comparison, reflecting floor effects in this relatively less salient driving problem. In contrast, cellphone distraction, aggressive driving, and road rage each produced four or more significant pairwise differences.

\subsection{Covariate Predictors of Class Membership}

\begin{figure}
\centering
\includegraphics[width=\columnwidth]{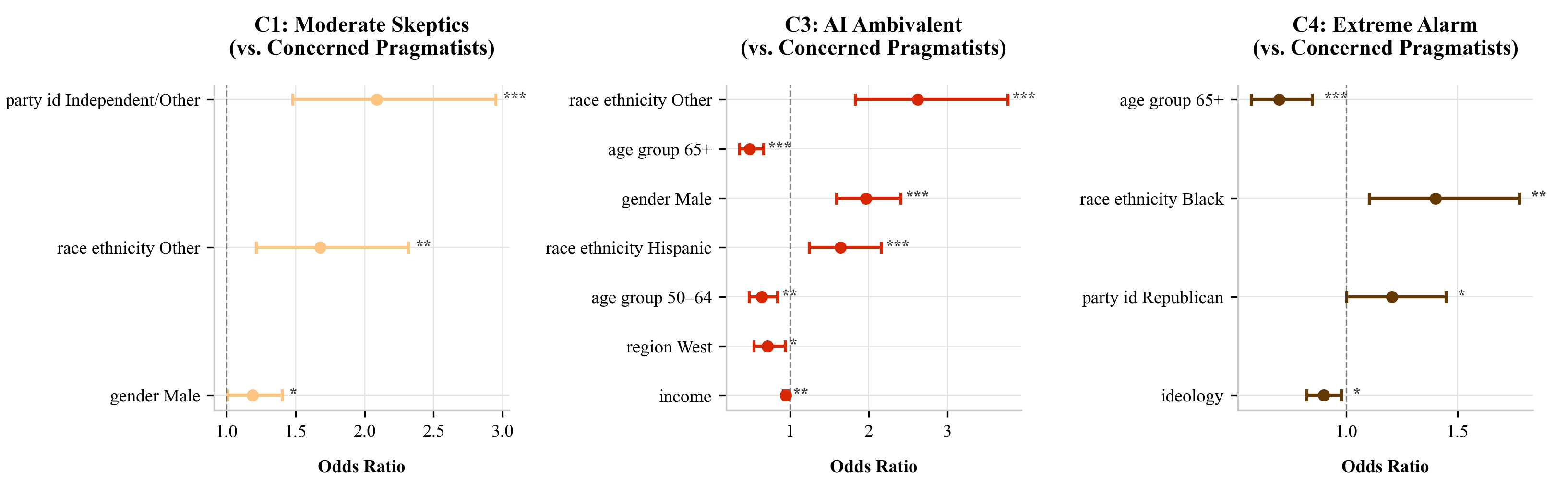}
\caption{Forest plot of odds ratios from the survey-weighted multinomial logistic regression, with Concerned Pragmatists (C2) as the reference class. Points indicate odds ratios; horizontal lines denote 95\% Wald confidence intervals. The dashed vertical line at OR $= 1.0$ marks no effect. Filled points indicate statistically significant coefficients ($p < 0.05$).}
\label{fig:or}
\end{figure}

Table~\ref{tab:or} and Figure~\ref{fig:or} present the survey-weighted multinomial logistic regression results, with Concerned Pragmatists (Class~2) serving as the reference category. The overall model achieved a McFadden pseudo-$R^2$ of 0.025, indicating that sociodemographic and ideological covariates explain a modest share of the variance in latent class membership. The Hausman--McFadden test confirmed the independence of irrelevant alternatives assumption (all $p = 1.0$), and all variance inflation factors remained below 1.80.

\begin{table}[H]
\caption{Multinomial logistic regression: odds ratios relative to Concerned Pragmatists (C2).}
\label{tab:or}
\centering
\begin{tabular}{@{} l c c c @{}}
\toprule
Covariate & C1: Mod.\ Skeptics & C3: AI Ambivalent & C4: Ext.\ Alarm \\
\midrule
Age 18--29 & 1.04 [0.83, 1.31] & 1.25 [0.97, 1.60] & 0.88 [0.72, 1.07] \\
Age 50--64 & 0.90 [0.72, 1.12] & 0.64 [0.48, 0.84]** & 1.00 [0.83, 1.19] \\
Age 65+ & 0.86 [0.68, 1.08] & 0.49 [0.36, 0.66]*** & 0.70 [0.57, 0.85]*** \\
Male & 1.19 [1.01, 1.40]* & 1.96 [1.59, 2.41]*** & 0.90 [0.78, 1.03] \\
Other gender & 0.41 [0.14, 1.17] & 0.89 [0.32, 2.46] & 0.38 [0.14, 1.00] \\
Education & 0.96 [0.90, 1.02] & 0.98 [0.91, 1.05] & 0.99 [0.95, 1.05] \\
Ideology (lib.\ to cons.) & 0.97 [0.87, 1.07] & 0.97 [0.86, 1.09] & 0.90 [0.82, 0.98]* \\
Republican & 0.81 [0.65, 1.01] & 1.02 [0.79, 1.32] & 1.20 [1.00, 1.45]* \\
Independent/Other & 2.09 [1.48, 2.95]*** & 1.30 [0.81, 2.08] & 0.79 [0.54, 1.17] \\
Rural & 1.04 [0.82, 1.32] & 0.80 [0.58, 1.10] & 0.87 [0.71, 1.07] \\
Income & 1.02 [0.99, 1.05] & 0.95 [0.91, 0.98]** & 0.99 [0.96, 1.01] \\
Black & 1.02 [0.76, 1.37] & 1.28 [0.91, 1.79] & 1.40 [1.10, 1.78]** \\
Hispanic & 1.23 [0.97, 1.57] & 1.64 [1.24, 2.16]*** & 1.02 [0.82, 1.26] \\
Asian & 0.97 [0.63, 1.49] & 1.42 [0.88, 2.28] & 1.26 [0.89, 1.77] \\
Other race & 1.68 [1.22, 2.32]** & 2.62 [1.83, 3.77]*** & 0.88 [0.64, 1.22] \\
Northeast & 0.87 [0.68, 1.10] & 0.86 [0.65, 1.15] & 1.01 [0.83, 1.24] \\
Midwest & 0.94 [0.74, 1.18] & 1.11 [0.85, 1.45] & 1.11 [0.92, 1.35] \\
West & 0.92 [0.74, 1.14] & 0.71 [0.54, 0.94]* & 1.07 [0.89, 1.29] \\
\bottomrule
\end{tabular}
\begin{flushleft}
\scriptsize\textit{Note.} Reference class: Concerned Pragmatists. Reference categories: Age 30--49, Female, Democrat/lean, Urban/Suburban, White, South. 95\% Wald CIs in brackets. Survey-weighted pseudo-MLE. $^{*}p < 0.05$, $^{**}p < 0.01$, $^{***}p < 0.001$. McFadden pseudo-$R^2 = 0.025$; all VIF $< 1.80$; IIA holds (all Hausman--McFadden $p = 1.0$).
\end{flushleft}
\end{table}

Gender was the strongest single predictor of AI Ambivalent membership: men were approximately twice as likely as women to belong to this least-concerned class (OR $= 1.96$, 95\% CI [1.59, 2.41], $p < 0.001$). Men were also marginally more likely to be Moderate Skeptics (OR $= 1.19$, $p < 0.05$). This gender gradient is consistent with the well-documented pattern of lower technological risk perception among men across multiple hazard domains \citep{Slovic1987}.

Age exhibited a selective rather than linear effect. Adults aged 65 and older were approximately half as likely as the 30--49 reference group to be AI Ambivalent (OR $= 0.49$, $p < 0.001$) and 30\% less likely to be Extreme Alarm (OR $= 0.70$, $p < 0.001$), indicating convergence toward the moderate center of the AI risk spectrum rather than occupancy of either extreme. The 50--64 age group showed a similar pattern for AI Ambivalent membership (OR $= 0.64$, $p < .01$). These findings indicate that the age effect on AI attitudes is more nuanced than the simple increase in concern reported in prior variable-centered analyses \citep{Tyson2023, Dreksler2025}: older adults are simultaneously less likely to occupy both the lowest-concern and the highest-concern classes.

Race and ethnicity predicted class membership in theoretically coherent ways. Black respondents were 40\% more likely than White respondents to be Extreme Alarm (OR $= 1.40$, $p < 0.01$), consistent with heightened AI risk perception among communities disproportionately affected by algorithmic systems in policing and credit scoring \citep{Bao2022}. Hispanic respondents were 64\% more likely to be AI Ambivalent (OR $= 1.64$, $p < 0.001$), a distinct pattern warranting further investigation. Political variables operated through two channels: party identification, rather than ideological self-placement, was the stronger predictor. Independents were more than twice as likely as Democrats to be Moderate Skeptics (OR $= 2.09$, $p < 0.001$), while Republicans were 20\% more likely to be Extreme Alarm (OR $= 1.20$, $p < 0.05$). Ideology showed a weak negative association with Extreme Alarm (OR $= 0.90$ per unit, $p < 0.05$), indicating that more liberal respondents were slightly more likely to express extreme AI concern. Education, urbanicity, and Census region were largely non-significant.

\subsection{Robustness Checks}

Table~\ref{tab:robust} summarizes the four robustness checks, all of which confirmed the stability of the four-class solution. Indicator sensitivity analysis demonstrated that the four-class structure was recoverable when any single indicator was omitted, with the maximum class prevalence shift of 0.085 falling below the 0.10 threshold. The BCH estimator produced results equivalent to the maximum-likelihood three-step estimator for continuous distal outcomes, while naive modal-assignment analysis exhibited mean attenuation of up to 0.064, confirming the importance of the classification error correction. Configural measurement invariance across Republican and Democrat subsamples was supported by Tucker's congruence coefficient $\phi = 0.96$, well above the 0.85 threshold for acceptable profile similarity \citep{LorenzoSevaTenBerge2006}. Random-start sensitivity analysis across 50 to 1,000 starting values yielded a log-likelihood range of exactly zero, confirming convergence to a unique global optimum.

\begin{table}[H]
\caption{Summary of robustness checks for the four-class solution.}
\label{tab:robust}
\centering
\begin{tabularx}{\textwidth}{@{} >{\raggedright\arraybackslash}p{3cm} >{\raggedright\arraybackslash}X c @{}}
\toprule
\textbf{Check} & \textbf{Result} & \textbf{Verdict} \\
\midrule
A. Indicator sensitivity & Max $|\Delta\pi| = 0.085 < 0.10$; four-class structure recovered in all nine leave-one-out models & Pass \\
\addlinespace
B. BCH vs.\ ML 3-step & BCH and ML three-step estimates equivalent for continuous outcomes; naive modal assignment exhibited attenuation up to 0.064 & Pass \\
\addlinespace
C. Measurement invariance & Tucker's $\phi = 0.96$ across Republican and Democrat subsamples (min $= 0.79$) & Pass \\
\addlinespace
D. Random starts & Log-likelihood range $= 0.000$ across 50 to 1,000 random starts & Pass \\
\bottomrule
\end{tabularx}
\end{table}

\section{Discussion}

The present study identified four empirically distinct latent classes of AI risk perception among U.S. adults and demonstrated that these classes are differentially associated with community driving-safety concerns, road rage frequency, and comparative evaluations of AI versus human driving performance. This section interprets the key findings, situates them within existing theoretical and empirical frameworks, and discusses implications for policy and communication.


The four-class solution, comprising Moderate Skeptics (17.5\%), Concerned Pragmatists (42.8\%), AI Ambivalent (10.6\%), and Extreme Alarm (29.1\%), provides the first person-centered taxonomy of AI risk perception derived from a large national probability sample. This taxonomy demonstrates that U.S. adults do not organize AI risk perceptions along a single continuum from low to high concern, as variable-centered analyses implicitly assume, but rather form qualitatively distinct subgroups that differ in both the intensity and the internal configuration of their concerns.

The finding of a dominant plurality (Concerned Pragmatists) endorsing ``very concerned'' responses without reaching extreme levels parallels the typology produced by \citet{Bao2022}, who used LCA on risk and benefit perception items in a national sample of 2,700 U.S. adults and identified five AI attitude segments: negative (33.3\%), ambivalent (28.5\%), tepid (24.0\%), ambiguous (7.5\%), and indifferent (6.6\%). The present study extends their work in three respects. First, our analysis draws on more recent data (2024 vs.\ 2020) and a larger probability-based sample ($n = 5{,}255$), capturing shifts in AI attitudes during a period of rapid generative AI deployment. Second, whereas \citet{Bao2022} segmented respondents on both risk and benefit perception items, our risk-focused indicator set provides a finer-grained decomposition of the concern dimension, distinguishing profiles such as Moderate Skeptics and Extreme Alarm that would fall within the broader ``negative'' category of their typology. Third, our study links AI risk perception profiles to an entirely separate behavioral domain, community driving-safety perceptions, rather than examining only within-domain AI attitudes. The absence of a ``purely positive'' segment in both studies is noteworthy: even our least-concerned class, the AI Ambivalent, displays genuine uncertainty rather than unqualified enthusiasm, suggesting that unconditional optimism about AI remains rare in the general U.S. population.

The four-class structure is also consistent with the broader audience segmentation literature. \citet{Leiserowitz2009} demonstrated that apparently normal distributions of climate change attitudes decompose into six discrete segments, each requiring distinct communication strategies. Our analysis confirms this principle in the AI risk domain: a single mean or regression coefficient cannot capture the qualitative differences between Moderate Skeptics (moderate concern, high distrust, high AI driving skepticism) and Extreme Alarm (maximum concern, maximum distrust, but intermediate AI driving evaluation). Fewer classes would merge meaningfully distinct profiles, while additional classes, as demonstrated by the nonviability of $K = 5$, fail to meet minimum size requirements for reliable estimation.


The most consequential empirical finding is the systematic association between AI risk perception profiles and community driving-safety perceptions. Respondents classified as Extreme Alarm perceived all six community driving hazards as more severe, reported driving-safety trends more pessimistically, and experienced road rage more frequently than members of any other class. Conversely, the AI Ambivalent class, which expressed the lowest AI concern, also perceived driving problems as least severe. This monotonic ordering is precisely what Cultural Theory of Risk predicts: risk perceptions covary across hazard domains because they are structured by underlying worldviews rather than domain-specific expertise \citep{Douglas1982}.

This cross-domain linkage also aligns with the psychometric paradigm's emphasis on shared cognitive appraisal dimensions. AI technologies and driving hazards differ in familiarity and voluntariness, yet both involve elements of controllability and dread that may activate common evaluative processes \citep{Slovic1987}. The finding that cellphone distraction exhibited the strongest class differentiation ($\chi^2(3) = 55.68$) is theoretically suggestive: among the assessed driving hazards, cellphone distraction is the most closely linked to personal technology use and may therefore tap the same technology-related risk sensitivity that underlies AI concern. By contrast, marijuana-impaired driving, which showed the weakest differentiation, involves a hazard with weaker ties to technology-related appraisals, consistent with the prediction that cross-domain covariance should be strongest for hazards sharing overlapping appraisal dimensions.

The AI versus human driving comparison produced a pattern that departs from the general severity ordering. Class~1 (Moderate Skeptics), not Class~4 (Extreme Alarm), expressed the most skeptical evaluation of AI driving relative to human driving. This result can be understood through the lens of AI trust rather than concern: Moderate Skeptics combine moderate concern with the second-highest distrust of AI for important decisions, whereas Extreme Alarm, though maximally concerned, may evaluate AI driving through the lens of general transportation pessimism rather than specific technology distrust. Class~3 (AI Ambivalent), which reports the highest AI trust, evaluated AI driving most favorably, further supporting the interpretation that trust, not concern level, is the proximal determinant of comparative driving performance evaluations. This distinction has direct implications for AV communication: messages aimed at influencing AI driving evaluations should target trust rather than general concern, as the two operate through partially independent psychological channels.


The covariate analysis revealed that sociodemographic characteristics explain a modest but theoretically informative share of class membership variance (McFadden $R^2 = 0.025$). This small effect size is itself a substantively important finding: it indicates that AI risk perception profiles are not primarily driven by demographics but instead reflect attitudinal orientations that cut across conventional social categories. This result echoes the finding of \citet{Bao2022}, who observed limited demographic differentiation across their five AI attitude segments and suggested that the relative absence of strong demographic sorting indicates a window of opportunity for public engagement before AI attitudes become entrenched along partisan or identity lines.

The gender effect was the strongest predictor, with men approximately twice as likely to occupy the AI Ambivalent class. This pattern is consistent with the well-documented gender gap in technological risk perception, in which men tend to judge technological hazards as less threatening \citep{Slovic1987}. Our person-centered analysis shows that this effect operates primarily by sorting men into the least-concerned class rather than uniformly reducing their concern across the distribution, a distinction that variable-centered methods cannot capture.

Age effects were selective rather than linear. Older adults (65+) were substantially less likely to be AI Ambivalent and moderately less likely to be Extreme Alarm, indicating convergence toward the moderate center of the AI risk spectrum. This finding partially diverges from prior research reporting that older adults are more concerned about AI \citep{Tyson2023, Dreksler2025}: the person-centered perspective reveals that older adults are simultaneously less likely to occupy both the lowest-concern and the highest-concern classes, a pattern consistent with moderation through greater life experience rather than simple age-related anxiety.

Racial and ethnic differences were consistent with expectations from the algorithmic fairness literature. Black respondents' elevated probability of Extreme Alarm membership aligns with heightened awareness of AI-related risks among communities disproportionately affected by algorithmic systems in policing, credit scoring, and healthcare \citep{Bao2022}. Hispanic respondents' higher probability of AI Ambivalent membership, a contrasting pattern, may reflect differential exposure to AI applications or distinct culturally mediated risk evaluation frameworks and warrants dedicated investigation.

The political findings merit attention in light of the observation by \citet{Bao2022} that AI attitudes in 2020 were ``not overtly politicized.'' Our 2024 data reveal nascent but detectable partisan structuring: Independents are strongly sorted into the Moderate Skeptics class (OR $= 2.09$, $p < 0.001$), suggesting that political disengagement or third-party orientation is associated with a distinctively moderate posture toward AI risks. Republicans are modestly more likely to express Extreme Alarm (OR $= 1.20$, $p < 0.05$), consistent with conservative skepticism toward novel technologies documented in AV acceptance research \citep{Mack2021, Peng2020}. That education, income, and urbanicity were largely non-significant reinforces the conclusion that AI risk profiles do not align neatly with standard socioeconomic stratification, a finding with direct implications for communication strategies that cannot rely on demographic targeting alone.


The four-class taxonomy offers actionable segmentation for AV communication and AI governance messaging. The Concerned Pragmatists (42.8\%), the largest class, represent the most strategically important audience. Their combination of serious but non-extreme concern with strong regulatory expectations suggests receptivity to transparency-focused messaging that acknowledges AI risks while detailing specific governance mechanisms \citep{Bao2025}. Because this class also perceives community driving hazards as moderately to highly severe, AV communication framed around safety improvements and crash reduction may resonate with their existing safety orientation.

The Extreme Alarm class (29.1\%) presents a distinct challenge. Messages that minimize AI risks or adopt a dismissive tone toward safety concerns are likely to be counterproductive for this audience, which is disproportionately composed of Black respondents and Republicans. Communication strategies emphasizing human oversight, accountability mechanisms, and incremental deployment may be more appropriate for engaging this segment, whose high driving-safety concern could paradoxically serve as a pathway to AV acceptance if framed in terms of replacing dangerous human driving behaviors.

The AI Ambivalent class (10.6\%), though small, is the only segment with a plurality expressing trust in AI and the most favorable evaluations of AI driving performance. This group represents potential early adopters and may serve as opinion intermediaries whose experiences could influence broader acceptance. The Moderate Skeptics (17.5\%) present a paradox for AV communication: their AI concern is moderate, yet they express the highest skepticism about AI driving relative to human driving. This disconnect suggests that trust, rather than overall risk perception, is the binding constraint, and communication for this audience should prioritize trust-building through transparent reporting of AV safety performance.

\section{Conclusion}

This study employed LCA on nationally representative survey data to identify empirically distinct profiles of AI risk perception among U.S. adults and to test whether these profiles are differentially associated with community driving-safety concerns. By integrating Cultural Theory of Risk, the psychometric paradigm, and audience segmentation theory within a unified person-centered design, the analysis addressed three gaps in the literature: the absence of a person-centered taxonomy of AI risk perception, the lack of empirical linkage between AI attitudes and driving-safety evaluations, and the near-total absence of BCH-corrected distal outcome methods in transportation research.

Three principal findings emerged. First, U.S. adults organize AI risk perceptions into four qualitatively distinct profiles rather than varying along a single continuum: Moderate Skeptics (17.5\%), Concerned Pragmatists (42.8\%), AI Ambivalent (10.6\%), and Extreme Alarm (29.1\%). Second, BCH-corrected analysis demonstrated that these profiles are systematically associated with all nine assessed driving-safety outcomes: respondents with higher AI concern perceive community driving hazards as more severe, report more pessimistic safety trends, and experience road rage more frequently. The exception, comparative evaluation of AI versus human driving, is driven by trust rather than concern level, with the most trusting class evaluating AI driving most favorably regardless of general concern. Third, sociodemographic covariates explain only a modest share of class membership variance, indicating that AI risk profiles cut across conventional demographic categories rather than mapping onto them, a finding with direct implications for communication strategies that cannot rely on demographic targeting alone.

These results carry two broader implications. For theory, the cross-domain covariation between AI risk profiles and driving-safety perceptions provides person-level evidence for the worldview-based risk structuring posited by Cultural Theory, extending this framework to the intersection of emerging technology and transportation safety. For practice, the four-class taxonomy offers a segmentation framework for AV communication: each class exhibits a distinct combination of concern intensity, trust orientation, and driving-safety assessment that warrants a tailored messaging strategy. As AI systems become more deeply integrated into transportation infrastructure, understanding the heterogeneous risk perceptions of the publics they are designed to serve will be essential for governance that is both empirically informed and democratically responsive.

Several limitations qualify these findings. The cross-sectional design precludes causal inference; the observed associations between AI risk profiles and driving-safety perceptions may reflect unmeasured third variables, and the direction of influence cannot be established. The analysis draws on a single survey wave (ATP Wave 152, August 2024) and is restricted to U.S. adults, so the four-class solution should be treated as a snapshot that may not generalize to other periods or national contexts. BVR analysis revealed 26 of 36 indicator-pair violations of local independence; although structurally interpretable and consistent with the documented behavior of polytomous ordinal LCA models, this pattern indicates residual within-class association that the four-class structure does not fully absorb. The modest covariate pseudo-$R^2$ (0.025) indicates that demographics capture only a small share of between-class variation. Finally, all measures are self-reported and were designed for general-purpose attitudinal monitoring rather than optimized for LCA indicator quality. Future research should address these gaps through longitudinal panel designs that can establish temporal stability and causal ordering, cross-national replication in European and East Asian settings with different AI governance and driving cultures, incorporation of psychographic variables (media consumption, personal AI experience, technology self-efficacy, institutional trust), and exploration of correlated-residual or hybrid LCA specifications that explicitly model within-class dependence among same-format items.

\printcredits

\bibliographystyle{cas-model2-names}

\bibliography{ref}




\end{document}